\documentstyle[12pt]{ioplppt}

\newcommand{\dii}{{\mathrm d}}
\newcommand{\ima}{{\mathrm i}}
\newcommand{\en}{{\mathcal N}}

\begin{document}
\jl{4}
\title{On the exact conservation laws in thermal models 
and the analysis of AGS and SIS experimental results}[On the exact 
conservation laws in thermal models]

\author{A.~Ker\"anen$^1$, J.~Cleymans$^2$ and E.~Suhonen$^1$}

\address{$^1$Department of Physical Sciences,
University of Oulu, FIN-90571 Oulu, Finland}

\address{$^2$Department of Physics, University of Cape Town, Rondebosch 7700,
South Africa}

\maketitle

\begin{abstract}
The production of hadrons in relativistic heavy ion collisions is
studied using a statistical ensemble with thermal and chemical
equilibrium. Special attention is given to exact conservation laws,
i.e. certain charges are treated canonically instead of using the usual
grand canonical approach.
For small systems, the exact conservation of baryon number,
strangeness and electric charge is to be taken into account. We have
derived compact, analytical expressions for particle abundances in
such ensemble. As an application, the change in $K/\pi$ ratios in AGS
experiments with different interaction system sizes is well
reproduced. 
The canonical treatment of three charges becomes
impractical very quickly with increasing system size.
Thus, we draw our attention to exact
conservation of strangeness, and treat baryon number and electric
charge grand canonically. We present expressions for particle
abundances in such ensemble as well, and apply them to reproduce the
large variety of particle ratios in GSI SIS 2 A GeV Ni--Ni experiments. 
At the energies considered here, the exact strangeness conservation 
fully accounts for strange particle suppression, and no extra
chemical factor is needed.
\end{abstract}

\section{Introduction}
The statistical models, based on the assumption of the quick natural entropy
maximization before the freeze-out in the relativistic heavy-ion
reaction, have been succesful in describing the total hadron abundancies
in various experiments, see
e.g. \cite{examples1}.  The usual method of grand
canonical ensemble is appropriate in case of very large interaction
systems. However, the systems studied experimentally are rather small,
so the exact conservation laws should be taken into account.
In such approach, the thermal analysis agrees well even with as
small systems as induced in CERN $e^{+}e^{-}$, $p-p$ and
$p-\overline{p}$ experiments \cite{examples2}. 

For larger systems, taking into account the
simultaneous conservation of baryon number $B$, strangeness $S$ and electric
charge $Q$ is more complicated. We review here the method developed,
and the application to AGS E802 $p-p$, $p-A$ and $A-A$ experiments in
the systematic manner to find if the increase in production of kaons
compared to pions can be due to finite volume effect incorporated in
relativistic canonical statistical mechanics \cite{cleymans2747}.   

If the net number of any relevant charge becomes large, the task
of computing abundancies in an ensemble respecting
exact conservation of $B$, $S$ and $Q$ becomes difficult. 
Letting $B$ and $Q$ fluctuate in the grand canonical manner, we may
focus to strangeness conservation. As an application, we have shown
that the particle ratios in GSI Ni+Ni experiments at 2 GeV A agree 
well with the statistical predictions \cite{cleymans3319}.

\section{The model}
In the usual grand canonical partition function, the states respecting
exact quantum number conservation are included, but only the
conservation of expectation values is demanded. 
By denoting the set of conserved quantum numbers by $\{C_i\}$, 
the canonical sum of
states, $Z_{\{C_i\}}$, can be projected from grand canonical one, $Z_G$ 
\cite{turko}. In case of $N$ $U(1)$ internal symmetries, the
projection takes the form
\begin{equation}
Z_{\{C_i\}} = \left[ \prod_{i=1}^{N} 
\frac{1}{2\pi}  \int_0^{2\pi}\dii\phi_i e^{-\ima C_i\phi_i}\right]
         Z_G(T,\{\lambda_{C_i}\}).
\label{eq:projection}
\end{equation}
Here we have assigned a group angle $\phi_i$ and a Wick -rotated 
fugacity factor $\lambda_i = e^{i\phi_i}$ for every exactly conserved
charge. 

\subsection{$Z_{B,S,Q}$}

Let us quote first the expression for the canonical partition function
respecting strict conservation of $B$, $S$ and $Q$.   
After putting in
the set of overall charges, $\{C_i\} = \{B,S,Q\}$, and performing some
algebraic excercise \cite{cleymans2747,cleymans_school}, 
we find
\begin{eqnarray}
Z_{B,Q,S}(T,V) &=& Z_0\left(\prod_{\nu=1}^{7}
\sum_{n_\nu=-\infty}^{\infty}\right) \nonumber \\
&\times&I_{-B+n_2+n_3+n_4+n_5+n_6+n_7}(2\en_n)\nonumber \\
&\times&I_{-Q+n_1+n_2-n_3+n_5-n_6+2n_7}(2\en_{\pi^c})  \\
&\times&I_{-S+n_1-n_4-n_5-n_6}(2\en_{K^0}) \nonumber \\
&\times&I_{n_1}(2\en_{K^c})I_{n_2}(2\en_{p})I_{n_3}(2\en_{\Delta^-})
\nonumber \\
&\times& I_{n_4}(2\en_{\Lambda})
I_{n_5}(2\en_{\Sigma^+})I_{n_6}(2\en_{\Sigma^-})I_{n_7}(2\en_{\Delta^{++}}).
\nonumber
\end{eqnarray}
Here we have used the modified Bessel functions $I_n$ with arguments 
$2\en_i$, where $\en_i$ is the sum of one particle partition functions
for particles carrying same set of quantum numbers, $\{B_i,S_i,Q_i\}$,
as the hadron $i$. $Z_0$ is the partition function for particles with 
vanishing quantum numbers considered.
In the expression above, we have omitted the
contribution from hadrons with $|S_i| > 1$, but the generalization is
straightforward \cite{cleymans2747}. 
The mean abundance of hadron $i$ in the system is 
\begin{equation}
\langle N_i \rangle = \frac{Z_{B-B_i,Q-Q_i,S-S_i}}{Z_{B,Q,S}} Z_i^1,
\end{equation}

The evaluation of the canonical partition function with three
simultaneously conserved quantum numbers becomes numerically very time
consuming for large values of $B$. So far, for systems with $B > 20$
we have been forced to resort to the grand canonical treatment.

\subsection{$Z_{S}$}

In large systems, such as Ni+Ni, the grand canonical treatment of
baryon number and electric charge is justified. However, the exact
strangeness conservation is still mandatory in theoretical considerations. 
If we only include the particles with $|S_i| \leq 1$, and require
vanishing net strangeness, we end up with \cite{cleymans3319,cleymans137}
\begin{equation}
Z_{S=0}(T,V,\lambda_B,\lambda_Q) = Z_0I_0(2\sqrt{\en_1\en_{-1}}),  
\label{eq:zs0}
\end{equation}
where the $N_{S_i}$ are the sums of grand canonical one-species partition
functions of particles carrying strangeness $S_i$ defined by
\begin{equation} \label{eq:Zi}
Z_i^1 = \lambda_B^{B_i}\lambda_Q^{Q_i}g_i 
\frac{V}{2\pi^2} \int_0^{\infty} \dii p\,  p^2
e^{\beta\sqrt{p^2+m_i^2}}.
\end{equation}
The equation (\ref{eq:zs0}) can be generalized to include the
particles with higher strangeness content as well, see
\cite{cleymans_school,cleymans137}.
Now the mean number of hadrons $i$ is
\begin{equation} \label{eq:mean1}
\langle N_i \rangle = Z_i^1
\left(\sqrt{\frac{\en_1}{\en_{-1}}}\right)^{S_i}
\frac{I_{S_i}(2\sqrt{\en_1\en_{-1}})}{I_0(2\sqrt{\en_1\en_{-1}})}.
\end{equation}

The nonlinear volume dependence of strange particle production rates
is visible in the $I_{S_i} / I_0$ coefficient in equation (\ref{eq:mean1}).
Whereas in grand canonical formalism the abundance of kaons is linear in
volume ($Z_i^1$), the $I_{1} / I_0$ term gives an additional
coefficient of first order in volume in small system limit.
This nonlinearity decreases smoothly with increasing volume, until in
thermodynamic limit it vanishes.    
  

\section{Applications and discussion}

\subsection{$K/\pi$ ratios in AGS experiment E802}
We have applied the $Z_{B,S,Q}$ in comparison with 
experimental $K/\pi$ production ratios reported by E802 collaboration,
see table \ref{tab:1}. 
\begin{table}
\caption{Experimental results reported by the E802 collaboration.
B and Q are calculated using geometrical considerations, see
\cite{cleymans2747} }
\label{tab:1}
\lineup
\begin{indented}
\item[]
\begin{tabular}{ccccccc}
\br
Collision & $K^{+}/\pi ^{+}$ & Ref. & $K^{-}/\pi ^{-}$ & Ref. & $B$ & $Q$ \\ 
\mr
\multicolumn{1}{l}{$p\   +\,  Be$} & 7.8$\pm $0.4\% & \multicolumn{1}{l}{
\cite{abb1,abb2}} & 2.0$\pm $0.2\% & \cite{abb1} & 3.9 & 2.3 \\ 
\multicolumn{1}{l}{$p\   +\, Al$} & 9.9$\pm $0.5\% & 
\multicolumn{1}{l}{\cite{abb2}} &  &  & 5.4 & 3.1 \\ 
\multicolumn{1}{l}{$p\   +\, Cu$} & 10.8$\pm $0.6\% & 
\multicolumn{1}{l}{\cite{abb2}} &  &  & 6.9 & 3.7 \\ 
\multicolumn{1}{l}{$p\   +\, Au$} & 12.5$\pm $0.6\% & 
\multicolumn{1}{l}{\cite{abb1,abb2}} & 2.8$\pm $0.3\% & \cite{abb1} & 9.7
& 4.5 \\ 
\multicolumn{1}{l}{$Si+Au$} & 18.2$\pm $0.9\% & 
\multicolumn{1}{l}{\cite{abb1}} & 3.2$\pm $0.3\% & \cite{abb1} & 102.7 & 
44.0 \\ 
& 19.2$\pm $3\% & \multicolumn{1}{l}{\cite{abb3}} & 3.6$\pm $0.8\% &
\cite{abb3} &  &  \\
\br
\end{tabular}
\end{indented}
\end{table}
In figure \ref{fig:1} we find the theoretical
curves for $K^+/\pi^+$ and $K^-/\pi^-$ ratios in constant baryon
density as functions of baryon number to be consistent with
experimental results. This shows that the increase in $K/\pi$ ratios
along with increasing system size can also be explained without any phase
transition or the in-medium effect in kaon masses \cite{schaffner,brown,weise}.

\subsection{Production ratios in GSI Ni+Ni experiments at 2 GeV A}

As an application of $Z_S$ the thermal model analysis is compared to 
experimental hadronic ratios in GSI Ni+Ni 2 GeV A experiments.
In this case, the widths of resonances affect substantially the
results, so we have applied the relativistic Breit-Wigner resonance
shape in computing the phase space integrals (\ref{eq:Zi}).

In Ni+Ni system, the isospin asymmetry ($\frac{B}{2Q} = 1.04$) has to be
taken into account by introducing chemical potential for electric
charge. This parameter, however, is eliminated by the simple
binding condition
between baryon- and charge densities, $n_B$ and $n_Q$: 
\begin{equation}
n_B = 2(\frac{B}{2Q})n_Q.
\end{equation}

The comparison between model predictions and experimental data is
collected in table \ref{tab:2}. 
\begin{table}
\caption{Particle ratios resulting from $Z_S$ compared to 
GSI Ni+Ni 2 GeV A experimental results. The best fit value,
$\mu_B = 0.72$ GeV, for the baryon chemical potential is used.}
\label{tab:2}
\begin{indented}
\item[]
\begin{tabular}{l|llll|lll}
\br
Ratio   & Model    &    &          &    & Data          & \\
R [fm]        & $4.2$ &  & $3$ &    &               & \\
T [MeV]       & $65$ &$75$ &$65$ &$75$ & ratio& 
        ref. \\ \mr 
${\rm K}^+/{\rm K}^-$  &$25.7$&$22.4$&$23.9$&$21.1$& $21\pm 9$
        &\cite{KaoS,oeschler,FOPI}\\ 
${\rm K}^+/\pi^+$  &$0.0071$&$0.0339$&$0.0027$&$0.0132$& $0.0074\pm
        0.0021$&\cite{oeschler,best,muentz}\\
$\phi/{\rm K}^-$   &$0.103$&$0.082$&$0.276$&$0.212$& $0.1$
     &\cite{herrmann} \\
$\pi^+/\pi^-$      &$0.893$&$0.895$&$0.894$&$0.898$& $0.89$    &\cite{pelte} \\
$\eta/\pi^0$    &$0.008$&$0.015$&$0.008$&$0.015$& $0.037\pm 0.002$ &
\cite{TAPS}\\
$\pi^+/{\rm   p}$  &$0.225$&$0.247$&$0.224$&$0.246$&  $0.195\pm
0.020$&\cite{muentz,pelte}\\
$\pi^0/{\rm B}$    &$0.104$&$0.108$&$0.104$&$0.107$& $0.125\pm 0.007$
        &\cite{TAPS} \\
${\rm d}/{\rm p}$  &$0.129$&$0.188$&$0.129$&$0.188$& $0.26$    &\cite{muentz} 
\\ \br
\end{tabular}
\end{indented}
\end{table}
Only the ratio $\eta/\pi^0$ does not
fit reasonably in thermal model scheme, but it serves as a clear indication
of incompleteness of the thermalized ideal gas model considered.
Apart from that, the other measured hadronic ratios
 are well described by the model. As a
clarification, the equal value curves for ratios in $(T,\mu_B)$ plane
in constant volume are shown in figure \ref{fig:2}.



\Figures

\begin{figure}
\caption{
Thermal model expectations for the production ratios $K^{+}/\pi ^{+}$
and $K^{-}/\pi ^{-}$
at a temperature of 100 MeV and a baryon density of 0.04 fm$^{-3}$ compared
to experimental results from the Brookhaven AGS. The experimental
ratios from $Si-Au$ collisions ($B\sim 103$) is moved to $B=21$ for the sake
of convenience.
}
\label{fig:1}
\end{figure}

\begin{figure}
\caption{Curves  in  the
$(\mu_B,T)$ plane corresponding to the GSI Ni+Ni 2 A GeV hadronic ratios
indicated.
The interaction volume corresponds to the radius of 4.2 fm, and the 
isospin asymmetry
is $B/2Q = 1.04$.}
\label{fig:2}
\end{figure}


\section*{References}

\begin{thebibliography}{99}

\bibitem{examples1} J. Sollfrank, \jpg  23 (1997) 1903

\bibitem{examples2}  F. Becattini, \ZP  C69 (1996) 485;
F. Becattini and U.~Heinz, \ZP  C76 (1997) 269

\bibitem{cleymans2747}
J. Cleymans, A. Ker\"anen, M. Marais and E. Suhonen, 
\PR  C56 (1997) 2474

\bibitem{cleymans3319}
J. Cleymans, D. Elliott, A. Ker\"anen and E. Suhonen, 
\PR  C57 (1998) 3319

\bibitem{turko}  K.  Redlich  and  L. Turko, \ZP  C5 (1980)
201;  L. Turko, \PL  B104 (1981) 153; R. Hagedorn and K.
Redlich, \ZP  C27 (1985) 541

\bibitem{cleymans_school}
J. Cleymans, A. Ker\"anen and E. Suhonen, presented at  the
11th Chris Engelbrecht Summer School in Theoretical Physics,
``Hadrosynthesis and Hadrons in Dense Matter'',  
Cape Town 4-13 February 1998, (to be published)

\bibitem{cleymans137}
J. Cleymans, K. Redlich and E. Suhonen, \ZP  C51 (1991) 137

\bibitem{abb1}  T. Abbot et al. (E802 Collaboration), \PRL
 66, (1991) 1567

\bibitem{abb2}  T. Abbot et al. (E802 Collaboration), \PR
 D45, (1992) 3906

\bibitem{abb3}  T. Abbot et al. (E802 Collaboration), \PRL 
 64, (1990) 847

\bibitem{schaffner}J. Schaffner, J. Bondorf and I. Mishustin,
\NP A625 (1997) 325
%
%
\bibitem{brown}Q.G. Li, C.H. Lee and G.E. Brown, 
\NP A625 (1997) 372
%
\bibitem{weise}  T. Waas, N. Kaiser, W. Weise, \PL B379
34; T. Waas, M. Rho and W. Weise, \NP A617 (1997) 44

\bibitem{KaoS}KaoS  Collaboration,  P.  Barth et al., \PRL
78 (1997) 4007
%
\bibitem{oeschler}KaoS  Collaboration,  H.  Oeschler,  presented at  the
11th Chris Engelbrecht Summer School in Theoretical Physics,
``Hadrosynthesis and Hadrons in Dense Matter'', Cape Town 4-13 February 1998, 
(to be published)
%
\bibitem{FOPI}FOPI Collaboration, 
Y. Leifels : ``FOPI Results - Strangeness in 4$\pi$.''
Talk  presented  at  the  International Workshop : ``Hadrons in
Dense Matter.'', GSI, Darmstadt, July 2-4, 1997

\bibitem{TAPS}  TAPS  Collaboration, 
M. Appenheimer et al., GSI 97-1, page 58; 
R. Averbeck, ``Hadronische
Materie  bei  SIS-Energien  -  Eine Thermodynamische Analyse.''
(unpublished),    presented   at   the   DPG   Fr\"uhjahrstagung,
G\"ottingen, February 1997

\bibitem{best} FOPI Collaboration,  D. Best et al., \NP A625
(1997) 367

\bibitem{herrmann} FOPI Collaboration, N. Herrmann, \NP A610
 (1996) 49c

\bibitem{pelte} FOPI Collaboration, D. Pelte et al. , 
\ZP A359 (1997) 55 

\bibitem{muentz} KaoS Collaboration, C. M\"untz et al. , 
\ZP A352 (1995) 175; \ZP A357 (1997) 399 

\end{thebibliography}
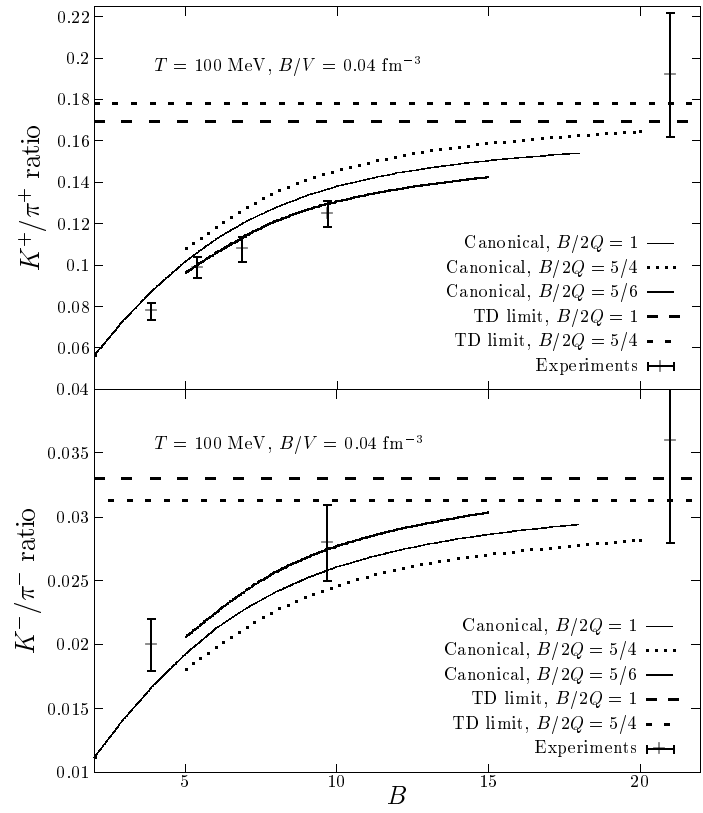
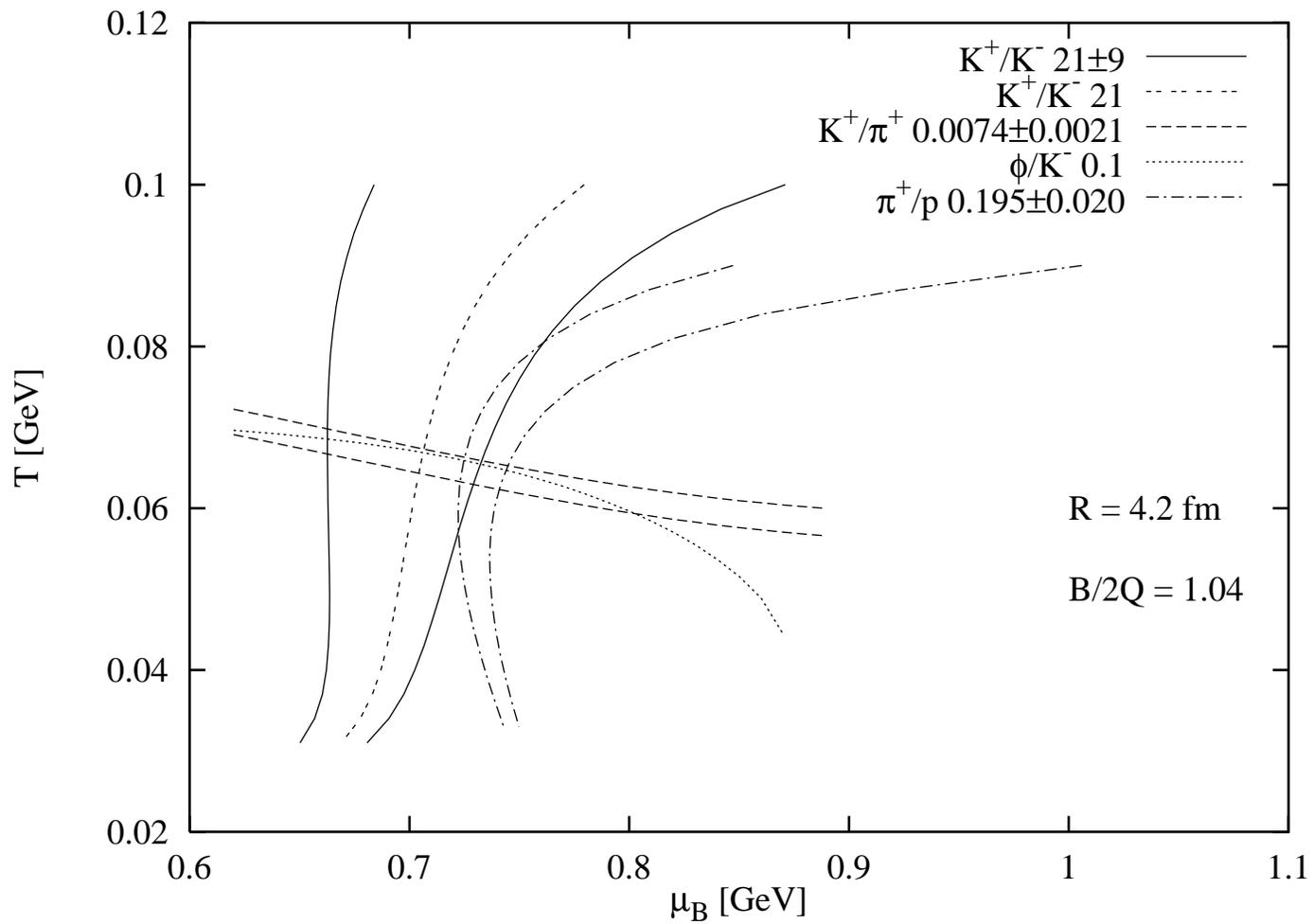
\end{document}